\documentclass{phbauth}        % Don't modify

\begin{document}
\def\beq{\begin{equation}}
\def\eeq{\end{equation}}

\begin{frontmatter}

\title{
Dzyaloshinskii-Moriya interaction in the paramagnetic state
and the polarized neutron scattering.}

\author%[address1]
{D.N.\ Aristov\thanksref{thank1}},
\author%[address1]
{S.V.\ Maleyev }

%\author{ D.N. Aristov, S.V. Maleyev \\

\address%[address1]
{Petersburg Nuclear Physics Institute, Gatchina
188300, Russia }

\thanks[thank1]{Corresponding author. E-mail: aristov@thd.pnpi.spb.ru}

\begin{abstract}
Dzyaloshinskii-Moriya (DM) interaction in the
paramagnetic state leads to the incommensurate spin fluctuations with
incommensurate vector proportional to the relative strength of the DM
interaction. We show that the DM interaction leads to helical spin
fluctuations which may be observed by the polarized neutron scattering.
\end{abstract}


\end{frontmatter}

In the case of inelastic magnetic scattering of polarized
neutrons the cross section consists of two terms. The first one
is independent on the initial neutron polarization ${\bf P}_0$
and determined by the symmetric part of the generalized magnetic
susceptibility $\chi_{\alpha\beta}(Q,\omega)$. The second one is
proportional to ${\bf P}_0$ and connected with the antisymmetric
part of $\chi_{\alpha\beta}$.  This antisymmetric part of
the susceptibility appears if the system is characterized by an
axial vector. There are two possibilities. i)~External magnetic
field or the sample magnetization (see \cite{1,2} and references
therein). ii)~Some intrinsic axial- vector interaction which is
connected to the noncentrosymmetry of the system \cite{3}.

In this paper we consider the case of the Dzyaloshinskii--Moriya
interaction (DMI) \cite{4} and demonstrate that the
dependence of the magnetic scattering
on ${\bf P}_0$
may appear in the
paramagnetic phase along with the incommensurate peaks in both
parts of the scattering cross section. We demonstrate it using
the DMI as perturbation in the three-dimensional (3D) case. Then we
confirm these results by exact solution of the $1D$ problem. It should
be noted here that the incommensurate ${\bf P}_0$-dependent
paramagnetic scattering was observed in MnSi \cite{5}. To the best of
our knowledge it is the only experimental study of this problem.

The DMI has the following form \cite{4}

       \beq
       V_{DM}\ =\
       \frac12\sum_{l,m}\ {\bf D}_{lm}[{\bf S}_l\times{\bf S}_m]\ ,
       \label{1}
       \eeq
where ${\bf D}_{lm}=-{\bf D}_{ml}$ is the DM axial vector. We
begin with the case when ${\bf D}_{lm}$ is invariant under
translations on the lattice, ${\bf D}_{lm} ={\bf D}_{{\bf
l}+{\bf a}_1,{\bf m} +{\bf a}_2}$, and
assume that ${\bf D}$ is directed along the $z$
axis.  After Fourier transform Eq.(\ref{1}) may be
represented as follows

       \beq
       V_{DM}\ =\
       i\sum_qd_{\bf q}S^x_{\bf q}S^y_{-\bf q}\ ,
       \eeq
where
       \beq
       d_{\bf q}\ =\
       -d_{-\bf q}=\ -i\sum_l D_{lm}e^{i{\bf qR}_{lm}}\ .
       \eeq

We assume now that the paramagnetic spin fluctuations are
isotropic, if one neglects the DMI.
In this case the spin Green
function has the form $G^0_{\alpha\beta}(q,\omega) =
\delta_{\alpha\beta}G^0(q,\omega)$.

Using interaction (2) as small perturbation we obtain
        \begin{eqnarray}
        G_{xx} &=& G^0+iG^0d_qG_{yx} \\
        G_{yx} &=& -iG^0 d_q G_{xx}\ . \nonumber
        \end{eqnarray}
As a result we get
        \begin{eqnarray}
        G_{xx}&=&
	G_{yy}=\ G^0\left(1-d^2_q(G^0)^2\right)^{-1}
     \\
	G_{xy} &=& -G_{yx}=\
	iG^0 d_q\left(1-d^2_q (G^0)^2\right)^{-1}.
	\nonumber
	\end{eqnarray}
We see that the DMI leads to the nondiagonal antisymmetric
components of the spin Green function. To clarify these
expressions let us consider the static approximation
$(\omega=0)$ and choose $G(q,0)$ in the conventional
Ornstein--Zernike form

        \beq
        G(q,0)\ =\ G(q)\ =\ A(q^2+\kappa^2)^{-1}\ ,
	\eeq
where $\kappa$ is the inverse correlation length and
$A\sim(T_ca^2)^{-1}$ where $T_c$ is the transition temperature
to the ordered state and $a$ is the interatomic spacing. Having
in mind that $d_0=0$ and at small $q$ one has $A d_q=2\alpha({\bf
q}\hat n)$ where $\hat n$ is the direction of the bonds, along which
the DM  interaction is present,
and $\alpha \sim ADa\ll
a^{-1}$ , we obtain from Eqs.\ (5) and (6)

     \begin{eqnarray}
     G_{xx}&=&
     G_{yy}=\frac A2
     \left(\left[\kappa_1^2+
     \left({\bf q}-{\alpha\hat n}\right)^2\right]^{-1}
     \right. \nonumber \\ && \left.
     +
     \left[ \kappa_1^2 +
     \left({\bf q}+{\alpha\hat n}\right)^2\right]^{-1}\right),
     \label{dmi-OZ} \\
     G_{xy}&=&
     -G_{yx} = i \frac A2
     \left(\left[\kappa_1^2 +\left({\bf q}+{\alpha\hat
     n}\right)^2 \right]^{-1}
     \right. \nonumber \\&& \left.
     - \left[\kappa_1^2+
     \left({\bf q}+{\alpha\hat n}\right)^2\right]^{-1}\right).
     \nonumber
     \end{eqnarray}

\noindent
with $\kappa_1^2=\kappa^2- \alpha^2$ ; these expressions describe
incommensurate spin fluctuations at ${\bf q}=\pm\alpha\hat n$.

These expressions are the result of the first order perturbation
theory in the DMI value and there should be additional terms of
order $\alpha^2$ in the denominators, which we did not evaluate.
According to Ref.\cite{6} due to the DMI the phase transition to
the ordered state should be the first order one. Experimental
study of this problem would be very interesting.
The possible candidates for such study could be the systems
MnSi, FeG, Fe$_2$O$_3$ and quasi-1D antiferromagnet CsCuCl$_3$.

As was stated above the antisymmetric part of the spin Green
function gives rise to the ${\bf P}_0$-dependent part of the
cross
section. In our case it may be represented as $G_{\alpha\beta} =
i\epsilon_{\alpha\beta\gamma}\widehat z_\gamma G^A$, where
$\widehat {\bf z}$ is the unit vector along the $z-$axis. In this
case the ${\bf P}_0$-dependent part of the cross section has the
form (cf.\cite{1})
     \begin{eqnarray}
     \left(\frac{d\sigma}{d\Omega d\omega}\right)_{P_0}
     &=& -\frac2\pi\ r^2f^2(q) \frac{k_f}{k_i}
     (1-e^{-\omega/T})^{-1}
     \nonumber \\ && \times
     ({\bf P}_0\hat {\bf q})(\hat {\bf q}\widehat {\bf z})
     \mbox{ Im }G^A({\bf q},\omega)\ ,
     \end{eqnarray}
where $r^2=0.292$ barns, $f(q)$ is the magnetic form-factor and
$\hat {\bf q}={\bf q}/q$.

If the asymmetry of $G_{\alpha\beta}$ is determined by the
magnetic field, Im$\,G^A$ is an even function of $\omega$ and,
provided $\omega\ll T$, we have $\int d\omega(d\sigma/d\Omega
d\omega)_{{\bf P}_0} =0$. It is a consequence of the $t$-oddness
of the magnetic field \cite{1,7}. In our case the vector $D$ is
$t$-even and for the static contribution one has

     \begin{eqnarray}
     \left(\frac{d\sigma}{d\Omega}\right)_{P_0} &=&
     \frac T\pi A r^2f^2(q)({\bf P}_0\hat {\bf q})
     (\hat {\bf q}\hat {\bf z})
     \left[
     \frac1{\kappa_1^2 +
     \left({\bf q}+{\alpha\hat n}\right)^2}
     \right.      \nonumber\\ &&
     \left.      -
     \frac1{\kappa_1^2 +
     \left({\bf q}-{\alpha\hat n} \right)^2}\right].
     \end{eqnarray}

Up to now we discussed the translationally invariant DMI. In the case
of the staggered DMI, one has $D_{l+b_x,m+b_y}=-D_{lm}$, where ${\bf
b}$ is the minimal vector along the bond where the DMI is present.
In this case,  instead of Eq.(2), we have

     \[
     V_{DM}\ =\ i\sum_q d_{\bf q}\,
     S^x_{\bf q} S^y_{-{\bf q}-{\bf k}_0}\ ,
     \]
where ${\bf k}_0$ is the AF reciprocal wave vector along ${\bf
b}$.  As a result $G_{xy}$ depends on ${\bf q}$ and ${\bf
q}+{\bf k}_0$ and cannot be determined in the neutron scattering
experiments. In this case the cross section is commensurate and
independent of ${\bf P}_0$.

The above results were obtained in the perturbation theory.
We present now an exact solution of the problem in the 1D case.
We consider the spin chain Hamiltonian of the form

        \begin{eqnarray}
        { H }  &=&
        \sum_{l=1}^{L}( J
        {\bf S}_l {\bf S}_{l+1}
        + {\bf D} [{\bf S}_l \times {\bf S}_{l+1} ])
        \label{iniHam}
        \end{eqnarray}

\noindent
with AF Heisenberg coupling $J>0$ and the Dzyaloshinskii-Moriya term
${\bf D}$. It is convenient to introduce here the quantity
$\delta = \tan^{-1} (D/J)$.

We observe that ${ H }$ is simplified upon
a canonical
transformation ${ H} \to e^{-iU} { H} e^{iU}$ with
        $        U = \delta \sum_{l=1}^L l\, S_l^z.        $
One can easily see that for the combinations $S_j^\pm = S_j^x \pm i
S_j^y$ we get
        \begin{equation}
        {\widetilde S}_l^{\pm} \equiv
        e^{-iU} S_l^\pm e^{iU} = S_l^{\pm}
        e^{\mp i l\delta }, \quad
        {\widetilde S}_l^z
        =S_l^z ,
        \label{newspins}
        \end{equation}

\noindent
and the Hamiltonian is reduced to the $XXZ$ model:

        \begin{equation}
        H = \sum_{l=1}^L (
         J_x
        ( {\widetilde S}^x_l {\widetilde S}^x_{l+1} +
        {\widetilde S}^y_l {\widetilde S}^y_{l+1})
        +  J {\widetilde S}_l^z {\widetilde S}_{l+1}^z )
        \label{newHam}
        \end{equation}

\noindent
with $J_x = J/\cos\delta$. It follows then, that the spectrum of the
initial problem (\ref{iniHam}) coincides with the one of
(\ref{newHam}). The observables in the initial system are recalculated
with the use of (\ref{newspins}) from the observables in the XXZ model
(\ref{newHam}). In the latter model one distinguishes the longitudinal
($G^\|(k,\omega)$) and the transverse ($G^\perp(k,\omega)$) spin
correlations, for the $z$ and $x$ components of spin, respectively.
The difference between these Green functions is small in the considered
limit, $\delta\to0$.

First we note that $U$ does not affect the $z-$component of
spins.  Therefore the ``longitudinal''  Green function
${G}_{zz}(k,\omega)= G^\|(k,\omega) $ has a commensurate
antiferromagnetic modulation.

The transverse spin susceptibilities look a bit
more complicated. Some calculation shows that

        \begin{eqnarray}
        G_{xx}(l,m,\omega) &=& G_{yy} (l,m,\omega)
        \nonumber \\ &=&
        G^\perp(l,m,\omega) \cos\delta(l-m)
        \\
        G_{yx}(l,m,\omega) &=& -G_{xy} (l,m,\omega)
        \nonumber \\ &=&
        G^\perp(l,m,\omega) \sin\delta(l-m)
        \end{eqnarray}

\noindent
In terms of the Fourier transform this reads as

        \begin{eqnarray}
        G_{xx}(q,\omega) &=& G_{yy} (q,\omega)
        \nonumber \\ &=& %   =
        \frac12[
        G^\perp(q+\delta,\omega)+
        G^\perp(q-\delta,\omega)],
        \nonumber
        \\
        G_{xy}(q,\omega) &=& -G_{yx} (q,\omega)
        \nonumber
        \\ &=& %   =
        \frac i2[
        G^\perp(q+\delta,\omega)-
        G^\perp(q-\delta,\omega)].
        \nonumber
        \end{eqnarray}

\noindent
From these expressions we see that the transverse and chiral
fluctuations are incommensurate along the chain and in the limit
$D/J\ll 1$ the incommensurate vector coincides with that determined by
Eq.(\ref{dmi-OZ}). However the complete solution of the problem
(\ref{newHam}) can be found in literature (see, e.g., Ref.
\cite{Schulz}).  In the 1D case we know the exact $\omega-$ and
$q-$dependence of all types of the spin fluctuations.

Note that in the quasi-1D compounds the value of $\delta\simeq D/J$,
determining the incommensurate wavevector of the fluctuations,
may be sufficiently large. For instance, one has
$\delta \simeq 0.18$ in the CsCuCl$_3$ and $\delta \approx
0.05$ in copper benzoate.  \cite{values} In the latter compound,
however, the presumably staggered variant of DMI should not lead to
consequences, observable by the polarized neutron scattering.

In conclusion, we demonstrate that the
DM interaction in the
paramagnetic state leads to the incommensurate spin fluctuations with
incommensurate vector proportional to the strength of the DM
interaction relative to the exchange one. It is shown also that DMI
leads to the helical spin fluctuations which may be observed by the
polarized neutron scattering.

This work was supported by Russian State Program for Statistical
Physics (Grant VIII-2), RFBR Grants No.\ 00-02-16873 and 00-15-96814,
and the Russian Program "Neutron Studies of Condensed Matter".

\end{document}